# Structural and electronic response upon hole-doping of rare-earth iron oxyarsenides $Nd_{1-x}Sr_xFeAsO$ ($0 < x \leq 0.2$)†

**Karolina Kasperkiewicz,[a] Jan-Willem G. Bos,[a] Andrew N. Fitch,[b] Kosmas Prassides,[c] and Serena Margadonna*[a]**

**Hole-doping of NdFeAsO via partial replacement of $Nd^{3+}$ by $Sr^{2+}$ is a successful route to obtain superconducting phases ($T_c$ = 13.5 K for a $Sr^{2+}$ content of 20%); however, the structural and electronic response with doping is different from and non-symmetric to that in the electron-doped side of the phase diagram.**

Upon electron doping, the family of rare-earth quaternary oxyarsenides with general formula REFeAsO (RE= rare earth) displays superconductivity with transition temperatures, $T_c$ as high as 55 K, being surpassed only by the high-$T_c$ cuprates.[1] The oxyarsenides adopt a tetragonal layered structure (space group *P*4/*nmm*) featuring alternating insulating RE-O and conducting Fe-As planes (Fig. 1a). On cooling, a structural phase transition to orthorhombic crystal symmetry (space group *Cmma*), accompanied by the development of long range AFM order occurs. Electron doping of the parent compounds REFeAsO (*e.g.* partial replacement of $O^{2-}$ by $F^-$ or oxygen deficiency) provides extra charge in the conduction plane, suppresses the structural/magnetic instability and triggers the occurrence of superconductivity. Recent theoretical calculations have shown that LaFeAsO is metallic (electron carriers) at the verge of a transition between a bad metal and a semiconductor[2] and that the $E_F$ is located at the edge of a peak and in a pseudogap region of the density of states (DOS), making the electronic structure strongly electron-hole asymmetric.[3,4] The Fermi surfaces of the REFeAsO parent compounds are characterised by the presence of both electron and hole pockets.[4] The family of Fe-based superconductors has now further expanded to include the related hole-doped superconductors, $A_{1-x}A'_xFe_2As_2$ (A= alkaline earth, A′= alkali metal)[5] as well as the ternary LiFeAs and the binary $FeSe_{1-x}$ phases.[6]

Hole-doping has been an extremely successful route for the preparation of high-$T_c$ superconductors in the $AFe_2As_2$ families. However, the issue of wheter hole doping in REFeAsO systems produces superconducting compositions is still open. There has been an early report on the substitution of $La^{3+}$ by $Ca^{2+}$ in LaFeAsO that does not induce superconductivity.[1] Subsequent work on $Sr^{2+}$ doping to afford

[a] *School of Chemistry, University of Edinburgh, Edinburgh, UK EH9 3JJ. E-mail: serena.margadonna@ed.ac.uk*
[b] *European Synchrotron Radiation Facility, 38043 Grenoble, France.*
[c] *Department of Chemistry, University of Durham, Durham, UK DH1 3LE.*
† Electronic Supplementary Information (ESI) available: results of the structural refinements at 5 and 300 K. See http://dx.doi.org/10.1039/b000000x/

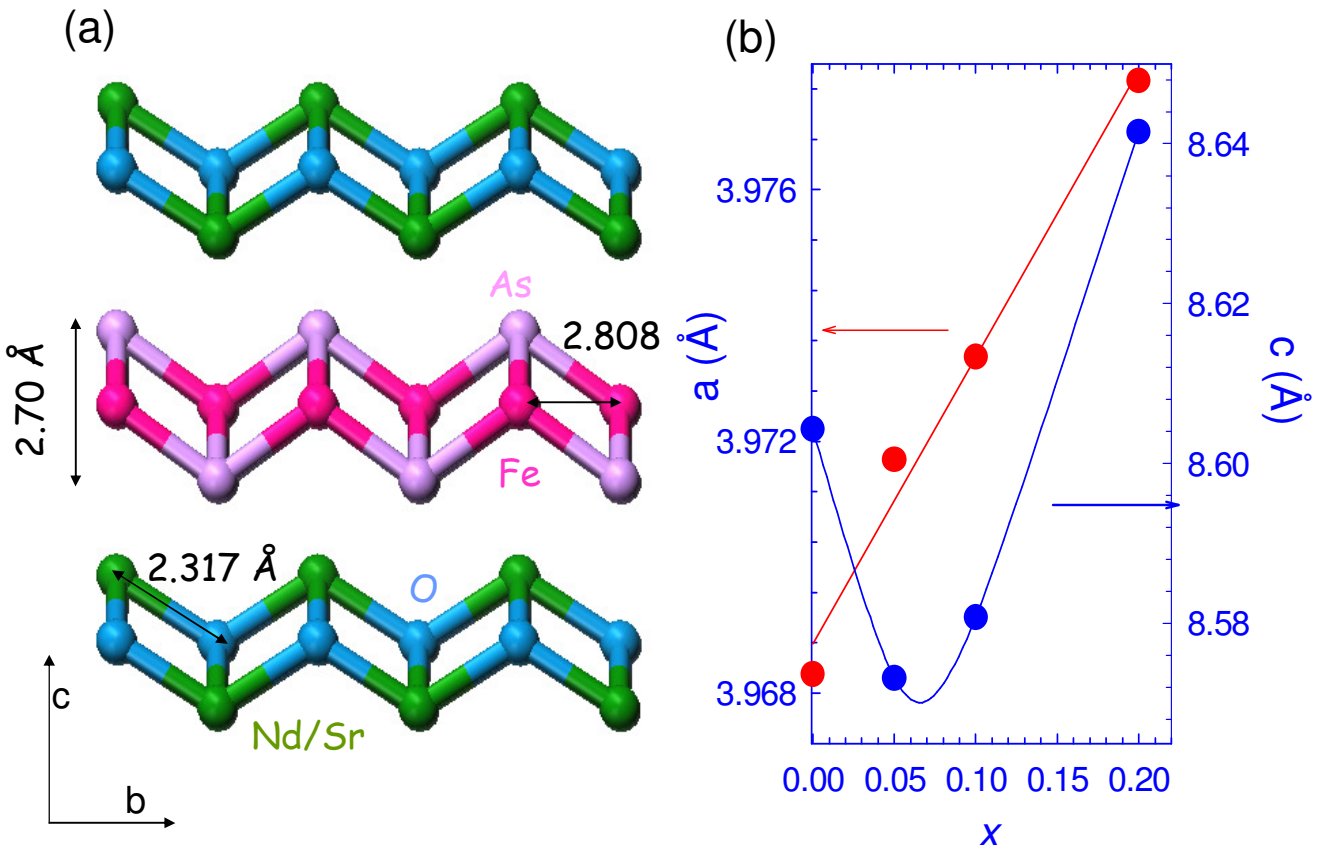

**Fig. 1 (a)** Schematic diagram of the room temperature tetragonal structure of $Nd_{1-x}Sr_xFeAsO$. The values of bond distances and angles reported refer to the composition with $x$ = 0.05. **(b)** Evolution of the room temperature tetragonal lattice constants as a function of $Sr^{2+}$ doping level $x$. The lines are guides to the eye.

$La_{1-x}Sr_xFeAsO$ demonstrated that superconducting phases can be achieved at doping levels $x > 0.08$ and that the evolution of $T_c$ shows a symmetric response to that on the electron-doped side. Positive Hall coefficients confirmed that the conduction is through hole charge carriers.[7] However, these results were not reproduced by other authors.[8] At this stage, it is essential to establish the role played by alkaline-earth-doping on the structural, electronic and magnetic behaviour of REFeAsO systems. Confirming the existence of superconducting phases, probing how $T_c$ varies with composition and revealing the influence of doping on the structural and magnetic transitions are of fundamental importance for the understanding of the pairing mechanism in these high-$T_c$ superconductors.

Here we report on the electronic and structural behaviour of a new family of hole-doped systems, $Nd_{1-x}Sr_xFeAsO$ ($x$ = 0.05, 0.1, 0.2). We find that introduction of $Sr^{2+}$ at the $Nd^{3+}$ site affords single phase materials adopting the *P4/nmm* structure at room temperature. Low temperature resistivity and susceptibility measurements reveal that compositions with small doping levels ($x$= 0.05) show semiconducting type behaviour. Increased doping leads to the emergence of metallic conductivity and superconductivity is observed at the maxium doping level, $x$ = 0.20 with $T_c$= 13.5 K. The results clearly demonstrate that hole doping of NdFeAsO can afford superconducting phases. In addition, the evolution of the electronic and structural properties with doping level is drastically different and non-symmetric to the electron-doped side of the phase diagram.

Powder samples of $Nd_{1-x}Sr_xFeAsO$ ($x$ = 0, 0.05, 0.1, 0.2) were prepared by reaction of stoichiometric quantities of high purity Fe, $Fe_2O_3$ and SrO with pre-synthesised FeAs and NdAs. All powders were mixed and grounded together and pressed into pellets in an Ar atmosphere glovebox and then heated for 48 hrs at 1100°C in evacuated quartz tubes. Increasing $x$ above 0.2 leads to samples containing secondary phases. The magnetic susceptibilities were measured using a Quantum Design MPMS magnetometer and the electrical resistivity was measured by the four-probe method with a Quantum Design PPMS instrument. High-resolution synchrotron X-ray diffraction experiments (λ= 0.40301 Å) were performed at various temperatures between 5 K and room temperature on the ID31 beamline at the European Synchrotron Radiation Facility (ESRF), France.[†]

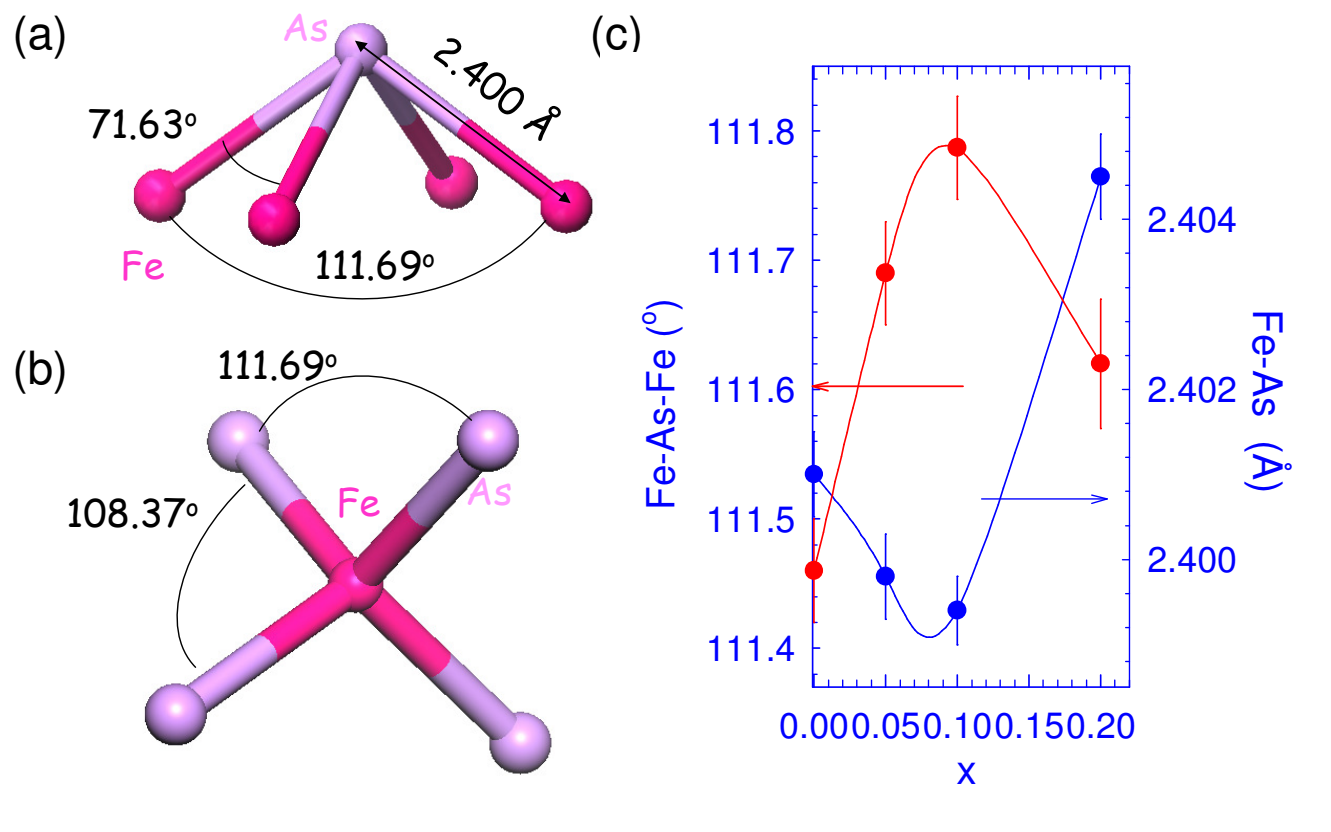

**Fig. 2** Schematic diagram of the **(a)** $AsF_4$ and **(b)** $FeAs_4$ units. The values of the bond distances and angles reported refer to the $x$= 0.05 composition. **(c)** Evolution of the room temperature Fe-As distance and of the Fe-As-Fe angle as a function of $Sr^{2+}$ doping level $x$. Lines are guides to the eye.

Inspection of the diffraction profiles of all $Nd_{1-x}Sr_xFeAsO$ compositions at room temperature readily reveals the tetragonal unit cell (*P*4/*nmm*) established before for the REFeAsO systems.[9,10] Rietveld analysis of the room temperature high statistics diffraction profiles confirmed the successful substitution of $Sr^{2+}$ at the $Nd^{3+}$ sites with the occupancies refining to 0.052(6), 0.091(6) and 0.21(1) for the nominal compositions $x$ = 0.05, 0.1 and 0.2, respectively (Table S1, Fig. S1). The structural analysis also revealed that with increasing doping level the $a$-axis increases monotonically as expected considering the larger ionic radius of $Sr^{2+}$ (1.18 Å *vs* 0.98 Å for $Nd^{3+}$) (Fig. 1b). $Sr^{2+}$ substitution occurs within the RE-O layers which determine the basal plane lattice dimensions. However, the $c$-axis shows a different response: it initially contracts at small doping levels before undergoing a considerable expansion at $x$> 0.1 (Fig. 1b). The $c$-axis dimensions are influenced by the geometry of the $FeAs_4$ tetrahedra (Fig. 2a,b) whose compression first decreases for 0< $x$ <0.1 and then increases at $x$> 0.1 (Fig. 2c). It is important to note that for small $Sr^{2+}$ content, the evolution of the structural parameters is exactly opposite to what is found for the electron doped systems where the $FeAs_4$ pyramidal units tend to compress monotonically with increasing doping level – the Fe-As-Fe angles decrease, while the thickness of the FeAs layer increases. However, at higher level of $Sr^{2+}$ substitution, the compression of the $FeAs_4$ tetrahedra increases in analogy with the trend observed for the superconducting electron doped phases (Fig. S2). It has been argued that the electronic structure of the REFeAsO systems strongly depends on small changes in interatomic distances and bond angles of the $FeAs_4$ pyramidal units. These structural parameters sensitively control both the Fe near- and next-near-neighbour exchange interactions as well as the width of the electronic conduction band due to the high degree of covalency (strong hybridisation) of the Fe-As bond.[11] The observed sudden increase of the Fe-As distance coupled with the larger lattice dimensions for the $x$= 0.2 composition may lead to a much smaller bandwidth and changes the electronic behaviour as compared to the small doping level compositions.

Diffraction profiles on compositions with $x$ = 0.05, 0.1 and 0.2 were collected on cooling between room temperature and 5 K. All samples showed a structural phase transition from tetragonal to orthorhombic (space group *Cmma*) in analogy with the undoped and electron-doped REFeAsO systems (Fig. 3 and S3, Table S2).[9,10] At the transition temperatures, no discontinuities are observed in either the $c$ lattice constant or the normalised volume, $V$. Increasing the $Sr^{2+}$ content level does not suppress the T→O phase transition and the transition temperatures, $T_s$ remain almost constant for all compositions ($T_s$ = 115 K for $x$ = 0.05 and 130 K for $x$ = 0.1 and 0.2, *cf.* $T_s$ = 135 K for NdFeAsO obtained in this work and in ref. 12).

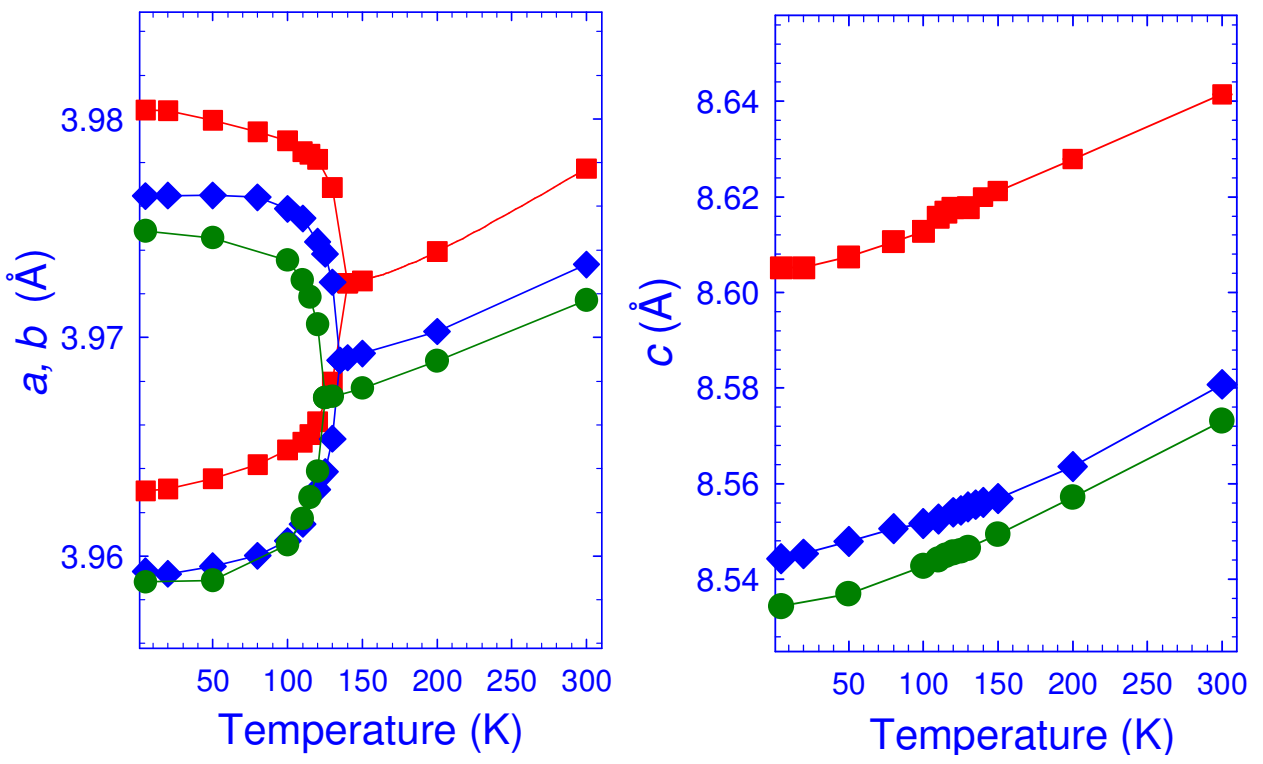

**Fig. 3** Temperature evolution of the lattice constants for $Nd_{1-x}Sr_xFeAsO$: green circles, blue diamonds and red squared refer to compositions with $x$ = 0.05, 0.1 and 0.2, respectively.

The zero-field-cooled magnetic susceptibilities, $\chi_M$ of $Nd_{1-x}Sr_xFeAsO$ for $x$= 0, 0.05, 0.1 and 0.2 were measured in an applied magnetic field of 1 T. The values of $\chi_M$ smoothly decrease as a function of doping level as expected from the substitution of $Nd^{3+}$ ($4f^3$, $S$=3/2) by the non-magnetic $Sr^{2+}$ in the RE-O layer ($\chi_M T$ values at 300 K: 1.66, 1.59 and 1.47 $cm^3$ K $mol^{-1}$ for $x$= 0.05, 0.1 and 0.2, respectively) (Fig. 4). As a first approximation, the $Fe^{2+}$ and $Nd^{3+}$ contribution to the magnetic susceptibility can be considered independently in the high temperature paramagnetic region (180< $T$ <300 K). It is then possible to subtract from the total magnetic susceptibility the paramagnetic $Nd^{3+}$ contribution by considering the ground state ($4f^3$, $^4I_{9/2}$) and a van Vleck contribution due to the first excited state ($^4H_{7/2}$). Such a subtraction leaves for NdFeAsO a temperature independent component which can be assigned to Pauli-like susceptibility of the $Fe^{2+}$ spins. At 300 K, the susceptibility after subtraction is $\chi_M = 8.3\times10^{-4}$ emu $mol^{-1}$ (*cf.* $\chi_M$ of LaFeAsO[10] with non-magnetic $La^{3+}$ is $5\times10^{-4}$ emu $mol^{-1}$).

Analogous treatment of the 1 T data of the other compositions performed by considering the nominal stoichiometry, showed that the values of the temperature independent component of the susceptibility first decreases for a Sr content of 5% ($9.3\times10^{-5}$ emu $mol^{-1}$) and then smoothly increases up to the maximum doping level of 20% ($1.95\times10^{-4}$ and $3.5\times10^{-4}$ emu $mol^{-1}$ for $x$ = 0.1 and 0.2, respectively). As the Pauli magnetic susceptibility is directly related to the density of states at the Fermi level, $N(E_F)$, the observed trend clearly indicates changes in the electronic properties upon increasing doping. Low-field (20 Oe) magnetic measurements showed that a superconducting transition is present only for the highest doping level composition, $x$ = 0.2 with a $T_c$ of 13.5 K and a superconducting fraction of 17% (Fig 4 inset).

Such trend was confirmed by resistivity measurements on samples with $x$ = 0, 0.05 and 0.2 (Fig. 4). The parent compound shows the typical behaviour already reported for REFeAsO.[1] The resistivity curve of the 5% doped material shows a drastic change with ρ increasing on cooling. It thus appears that upon low levels of hole doping, the electron charge carriers in NdFeAsO are initially depleted. This is consistent with the electron cylinders of the Fermi surface[4] decreasing in volume and this is not immediately compensated by an enlargement of the hole pockets[4] pushing the system through a transition from a bad metal to a semiconductor. However, further hole doping increases the amount of hole-type charge carries, inflates the volume of the hole pockets and metallic-type conductivity is restored.[‡] Indeed, the 20% doped system shows a metallic type of behaviour with ρ decreasing with temperature and a superconducting transition at $T_c$ =13.5 K in agreement with the magnetisation data. We also note that the derivative of the temperature-dependent resistivity, dρ/d$T$ shows maxima at 140, 115 and 130 K for $x$ = 0, 0.05 and 0.2. These correspond exactly to the temperatures at which the T→O structural transition occurs.

There are many important points arising from the present results. Partial replacement of $Nd^{3+}$ by $Sr^{2+}$ to afford $Nd_{1-x}Sr_xFeAsO$ leads to superconducting phases at doping levels with $x$>0.1. However, the electronic and structural behaviour upon hole doping is different and non-symmetric to the corresponding electron-doped systems where a smooth evolution of the structural and electronic properties is observed. In the hole doped systems, the T→O phase transition is not suppressed up to $x$~0.2 and in addition the temperature at which it occurs varies little in sharp contrast to the structural behaviour of the electron-doped analogues.[9] The evolution of the structural parameters of the $FeAs_4$ tetrahedra shows drastic changes at doping levels $x$>0.1 with a large increase in the Fe-As distances which is accompanied by a decrease in the thickness of the Fe-As layers. Magnetisation and resistivity measurements also show considerable differences in the electronic behaviour with increasing $x$. All our experimental observations describe a scenario where at small doping levels, the number of carriers available decreases, leading to an accompanying decrease in $N(E_F)$ and initially pushing the system into the semiconducting regime. Higher $Sr^{2+}$ content is necessary for the metallic regime to re-emerge. At $x$> 0.1, the weaker Fe-As hybridisation causes a decrease in the conduction bandwidth which is accompanied by an increase in $N(E_F)$ and the occurrence of superconducting phases. It appears that for NdFeAsO higher levels of hole doping are necessary to observe a similar phenomenology to the electron doped systems as an increased number of carriers is necessary to first overcome the initial semiconducting behaviour. This is in contrast to $NdFeAsO_{1-x}F_x$, in which there is a continuous change of the electronic properties upon F doping and superconductivity is obtained for $x$ >0.08 with much higher $T_c$ (*cf.* $NdFeAsO_{0.88}F_{0.12}$, $T_c$= 50 K).[13] It is likely that if it were chemically possible to obtain levels of $Sr^{2+}$ doping higher than 20%, a further increase in $T_c$ could be achieved. We note that in hole-doped $K_{1-x}Sr_xFe_2As_2$, superconductivity is also reported at doping levels of 20% ($x$= 0.80) with a $T_c$ of ~15 K.[5] Higher transition temperatures ($T_c$~37 K) are obtained by increasing the doping level to 40-50%.

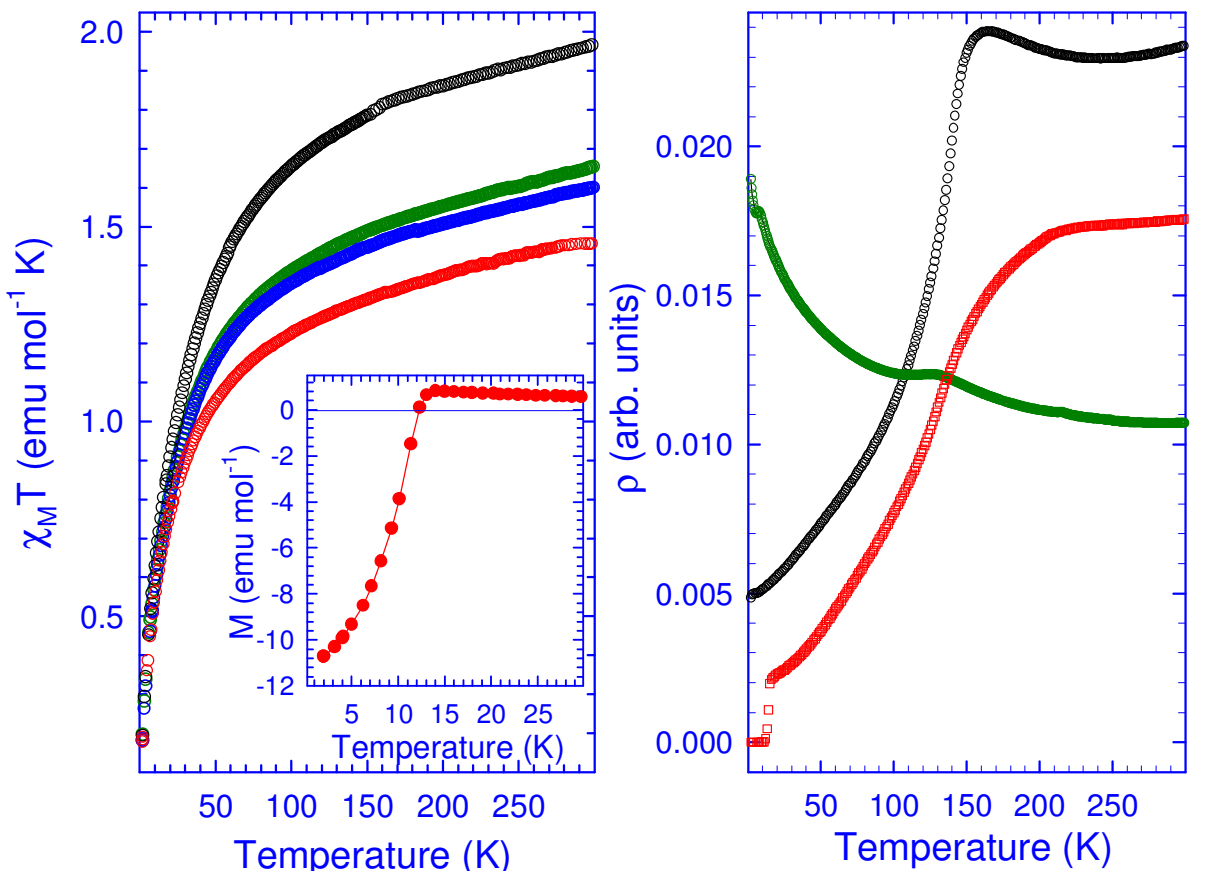

**Fig. 4** Left panel: temperature evolution of $\chi_M T$ for $Nd_{1-x}Sr_xFeAsO$: black, green, blue and red circles refer to compositions with $x$ = 0, 0.05, 0.1 and 0.2 respectively. Right panel: temperature evolution of the resistivity for $x$ = 0 (black), $x$ = 0.05 (green) and $x$ = 0.2 (red).

## Notes and references

† The structural analysis of NdFeAsO is not discussed here as it is in agreement with that reported in the literature.[12]

‡ Altough measurements of the Hall coefficients are necessary to determine what type of carriers are responsible for the conduction, it has been already demonstrated that $Sr^{2+}$ doping leads to hole type conduction in systems such as $La_{1-x}Sr_xFeAs$[10] and $Sr_{1-x}K_xFe_2As_2$[5].

1 Y. Kamihara, T. Watanabe, M. Hirano and H. Hosono, *J. Am. Chem. Soc.* 2008, **130**, 3296; X. H. Chen, T. Wu, G. Wu, R. H. Liu, H. Chen and D. F. Fang, *Nature* 2008, **453**, 761
2 K. Haule, J.H. Shim and G. Kotliar, *Phys. Rev. Lett.* 2008, **100,** 226402.
3 Z. P. Yin, S. Legebue, M. J. Han, B. Nea., S. Y. Savrasov and W. E. Pickett, arXiv:0804.3355, 2008.
4 D. J. Singh and M. H. Du, arXiv:0803.0429, 2008.
5 K. Sasmal, B. Lv, B. Lorenz, A. Guloy, F Chen, Y, Xuw and C. W. Chu, arXiv:0806.1301, 2008.
6 M. J. Pitcher, D. R. Parker, P. Adamson, S. J. C. Herkelrath, A. T. Boothroyd and S. J. Clarke, arXiv:0807.2228, 2008; S. Margadonna, Y. Takabayashi, M. T. McDonald, K. Kasperkiewicz, Y. Mizuguchi, Y. Takano, A.N. Fitch, E. Suard and K. Prassides, *Chem. Commun.* 2008, DOI:10.1039/ b813076k.
7 G. Mu., L Fang, H. Yang, X. Zhu, P. Cheng and H.H. Wen, *Europhys. Lett.* 2008, **82**, 17009 and arXiv: 0806.2104, 2008
8 G. Wu, H. Chen, Y. L. Xie, Y. J. Yan, T. Wu, R. H, Liu, X. F. Wang, D. F. Fang, J. J. Ying and X. H. Chen, arXiv: 0806.1687, 2008.
9 S. Margadonna, Y. Takabayashi, M.T. McDonald, M. Brunelli, G. Wu, R. H. Liu, X. H. Chen and K. Prassides, arXiv: 0806.3962, 2008.
10 T. Nomura, S.W. Kim, Y. Kamihara, M. Hirano, P. V. Susko, K. Kato, M. Takata, A. L. Shluger and H. Hosono, arXiv: 0804.3569, 2008.
11 V. Vildosola, L. Pourovskii, R. Arita, S. Biermann and A. Georges, arXiv: 0806.3285, 2008.

12 M. Fratini, R. Caivano, A. Puri, A. Ricci, Z.-A. Ren, X.-L. Dong, J. Yang, W. Lu, Z.-X. Zhao, L. Barba, G. Arrighetti, M. Polentarutti and A. Bianconi, *Supercon. Sci. Techno*l. 2008, **21**, 092002.
13 G. F. Chen, Z. Li, D. Wu, J. Dong, G. Li, W. Z. Hu, P. Zheng, J. L. Luo, and N. L. Wang, *Chin. Phys. Lett*. 2008, **25**, 2235.